\documentclass[12pt]{article}
\usepackage{natbib}
\usepackage{graphicx}
\begin{document}
\def\teff{$T\rm_{eff }$}
\def\kms{$\mathrm {km s}^{-1}$}

\title{
Concept study of a small Compton polarimeter to fly on a CubeSat
}

\author{
Yi-Chi Chang$^1$,
Chien-Ying Yang$^1$,
Hung-Hsiang Liang$^1$,
\\Che-Yen Chu$^1$,
Jeng-Lun Chiu$^2$,
Chih-Hsun Lin$^3$,
Philippe Laurent$^4$,
\\and Hsiang-Kuang Chang$^{1,5}$
\\
\\
\small{$^1$Institute of Astronomy, National Tsing Hua University, Hsinchu, Taiwan}
\\
\small{$^2$National Space Organization, National Applied Research Labs, Hsinchu, Taiwan}
\\
\small{$^3$Institute of Physics, Academia Sinica, Taipei, Taiwan}
\\
\small{$^4$CEA/DRF/IRFU/DAp, Saclay, France}
\\
\small{$^5$Department of Physics, National Tsing Hua University, Hsinchu, Taiwan}
\vspace{0.3cm}
\\
\small{{\it ycchungvis@gmail.com; hkchang@mx.nthu.edu.tw}}
\vspace{0.3cm}
\\
\small{In Proceedings of the 12th INTEGRAL conference and 1st AHEAD Gamma-ray Workshop,} 
\\
\small{Geneva (Switzerland), 11-15 February 2019, Ed.\ C. Ferrigno, E. Bozzo, P. von Ballmoos,}  
\\
\small{to appear in the Journal of the Italian Astronomical Society.}
}


\date{}
\maketitle

\abstract{
Application of cubesats in astronomical observations has been getting 
more and more mature in recent years. Here we report a concept study of
a small Compton polarimeter to fly on a cubesat for observing 
polarization of soft gamma-rays from a black-hole X-ray binary, Cygnus 
X-1. Polarization states provide very useful diagnostics on the 
emission mechanism and the origin of those gamma rays. In our study, we
conducted Monte Carlo simulations to decide the basic design of this 
small polarimeter. Silicon detectors and cerium bromide scintillators 
were employed in this study. We estimated its on-axis Compton efficiency at
different energies and its data telemetry requirement when flying in a 
low earth orbit. 
Our results indicate that it is feasible to achieve high signal-to-noise ratio for observing Cyg X-1 
with such a small instrument. Based on this study, we will proceed to have a more 
realistic design and look for opportunities of a cubesat space mission.}

\section{Introduction}

Astronomical space missions, with satellites weighing a few hundred kilograms and above, are very expensive. 
Their developing time is also very long. In recent years, cubesats of combined standardized units of 10 cm cubic each, so-called 1 U, 
have found more and more applications. Each unit weighs about one to two kilograms. 
Because of its light weight and standardized specification and supporting technology, 
the cost and developing cycle can be significantly reduced. In astronomy, 
several exciting cubesat missions have been conducted, being developed, or proposed (e.g. \citet{shkolnik18}).

In this paper we report a concept study of a small Compton polarimeter, suitable to fly on board a 3-U cubesat, 
for measuring the polarization of soft gamma-rays from Cyg X-1.
Although the instrument is small, a dedicated mission to observe a single target all the time when technically possible
may still achieve its science goal. Our study is to examine this possibility, in addition to determining a better model design.

Cyg X-1 is a galactic X-ray source and is believed to contain a black hole \citep{sunyaev79,gierlinski97}, 
which is estimated to be about 14.8 solar mass. 
Cyg X-1 is in a high mass X-ray binary (HMXB) system located at about 2 kpc from the Sun. Its companion is a blue supergiant variable, HDE 226868. 
The origin of the hard X-ray and soft gamma-ray emission, up to several hundred keV for typical accreting black-hole systems and up to or beyond MeV energies for the brighter members of this group, remains elusive after several decades of investigation.
The emission from the keV regime up to about 200 keV is
usually interpreted as being due to thermal emission from the accretion disk and Comptonization 
by a thermal distribution of hot electrons, while that at even higher energies calls for a new interpretation. 
X-ray binaries sometimes show relativistic jets, as seen in radio and IR bands, 
and the jet is possibly the origin of that high energy emission component. 
If that is the case, that emission is likely highly polarized.
Earlier INTEGRAL observations showed that the emission from Cyg X-1 above about 400 keV indeed has a high polarization degree
($67\pm 30$\% with IBIS \citep{laurent11} and larger than 75\% with SPI \citep{jourdain12}), 
and the levels of polarization are constrained to be lower than 20\% below 
400 keV (IBIS) and below 200 keV (SPI).  It will be very helpful to have long-term monitoring of Cyg X-1 to pin down the polarization state at
these energies and to see whether there is any polarization evolution between its low-hard and high-soft states.  

\section{Instrument Models}

\begin{figure*}
\resizebox{\hsize}{!}{\includegraphics[clip=true]{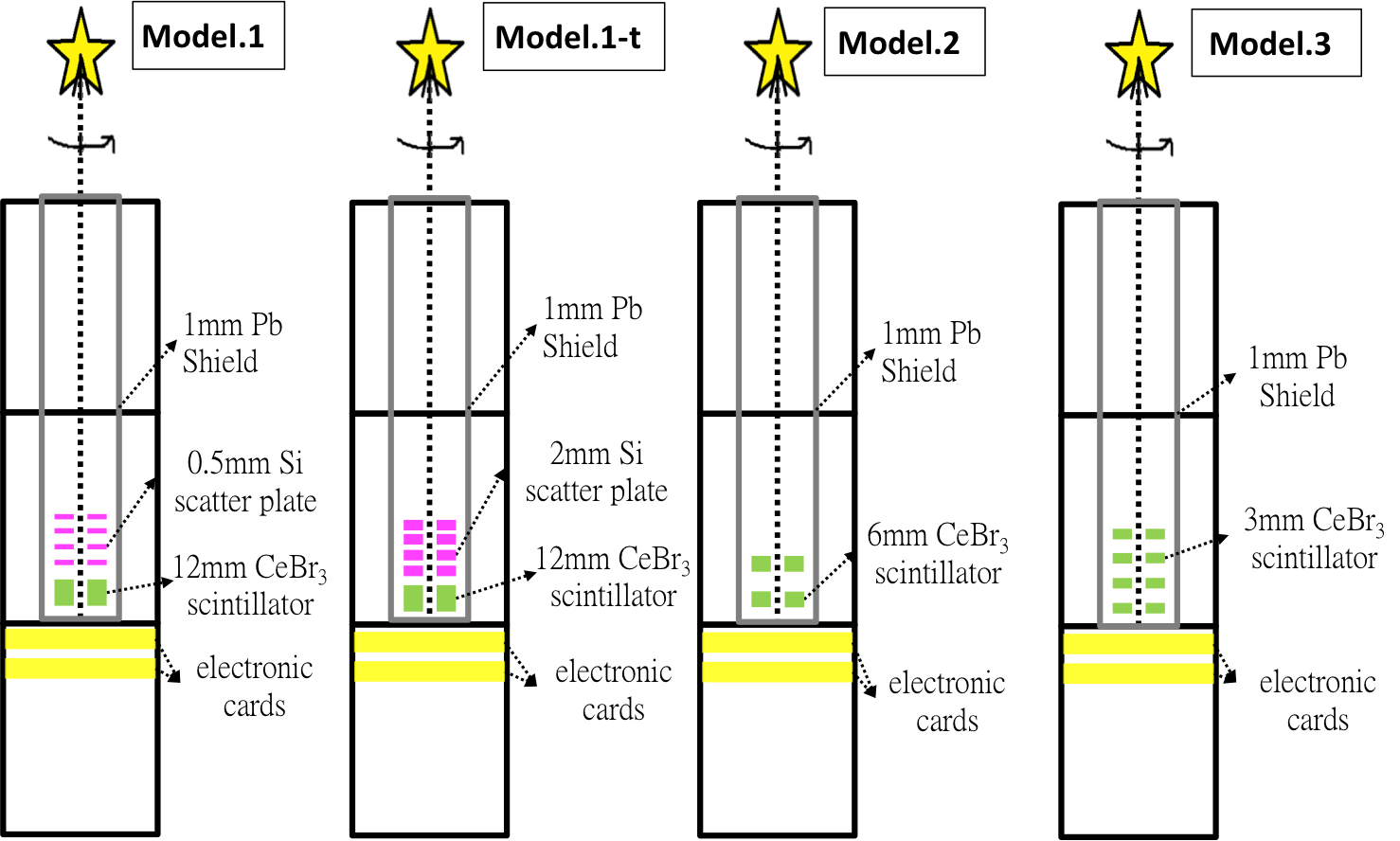}}
\caption{
The four instrument models studied in this paper. 
The numbers shown in this figure are all for thickness.  
See the main text for more descriptions.
The instrument with its shield is located in the upper 2 U of a 3-U cubesat.
When pointed towards Cyg X-1 for observation, a slow rotation around the pointing direction is desired
for eliminating systematic bias for the azimuthal scattering angle distribution, which is of essential importance in
polarization measurement.  
}
\label{model}
\end{figure*}
We started with the conventional idea underlying COMPTEL 
with a scatter module at the top and an absorber module at the bottom.
The method of time-of-flight, however, is not applicable because our instrument is small.
We employ Compton event reconstruction without electron tracking.
The first model (Model 1) we consider is shown in the left panel in Figure \ref{model}.
It consists of 4 layers of silicon sensors at the top and one layer of cerium bromide (CeBr$_3$) scintillator at the bottom.
Each layer of silicon sensors is an array of $2\times 2$ double-sided silicon strip detectors (DSSD). 
Each DSSD is of dimension $10\times 10\times 0.5$ mm$^3$ with 1-mm strip pitch. 
This strip pitch gives a spatial resolution similar to some silicon drift detectors (SDD) available in our lab, which we may
also consider to use. 
The cerium bromide layer at the bottom is an array of $2\times 2$ cerium bromide units, each  of $12\times 12\times 12$ mm$^3$
and wrapped with teflon on its top and 4 sides. SiPM with $4\times 4$ readout channels is attached to the bottom of each cerium bromide unit.
From the signal distribution among the 16 SiPM channels, it is possible to determine the interaction location to 3 mm uncertainty \citep{gostojic16}.

Model 1 as described above requires 160 readout channels from its silicon module and 64 channels from the cerium bromide one.
To increase the number of scattering in the silicon layers, we also defined another similar model but with the thickness of silicon sensors being 2 mm, instead of 0.5 mm.
This model is called Model 1-t, the second from the left in Figure \ref{model}.

We also defined another model, Model 2, to have thinner cerium bromide scintillator units, that is, each of $12\times 12\times 6$ mm$^3$,
and at the same time to use $2\times 2$ cerium bromide arrays of such units both at the top and at the bottom.  
SiPM arrays, similar to the one used for Model 1, are attached to the bottom of the two cerium bromide layers for readout.
Model 2 has altogether 128 readout channels.
We further defined Model 3 to have even thinner cerium bromide scintillators, that is,  each of $12\times 12\times 3$ mm$^3$,
and to have 4 layers of $2\times 2$ arrays of such units. SiPM arrays are attached to the bottom of all the 4 layers. 
Model 3 therefore has 256 readout channels.

The space between all the layers is 1 cm.
All the models are with a 20-cm long, 2.5-cm-squared-tube Pb shield of thickness 1 mm,
which shields the instrument on the surrounding four sides and on the bottom.   
This shielding was adopted based on the simulation results described in the next section.

\section{Performance Simulations}

We conducted Monte Carlo simulations to study the Compton efficiency of each model, to 
examine effects of different shielding models, to estimate the count rate of photons from Cyg X-1, 
and to estimate the data rate (mainly due to background at LEO) for transmission to the ground stations.
The simulation tool employed in our study is  the `Medium Energy Gamma-ray
Astronomy library' (MEGAlib) \citep{zoglauer08}.

\begin{figure*}[]
\resizebox{\hsize}{!}{\includegraphics[clip=true]{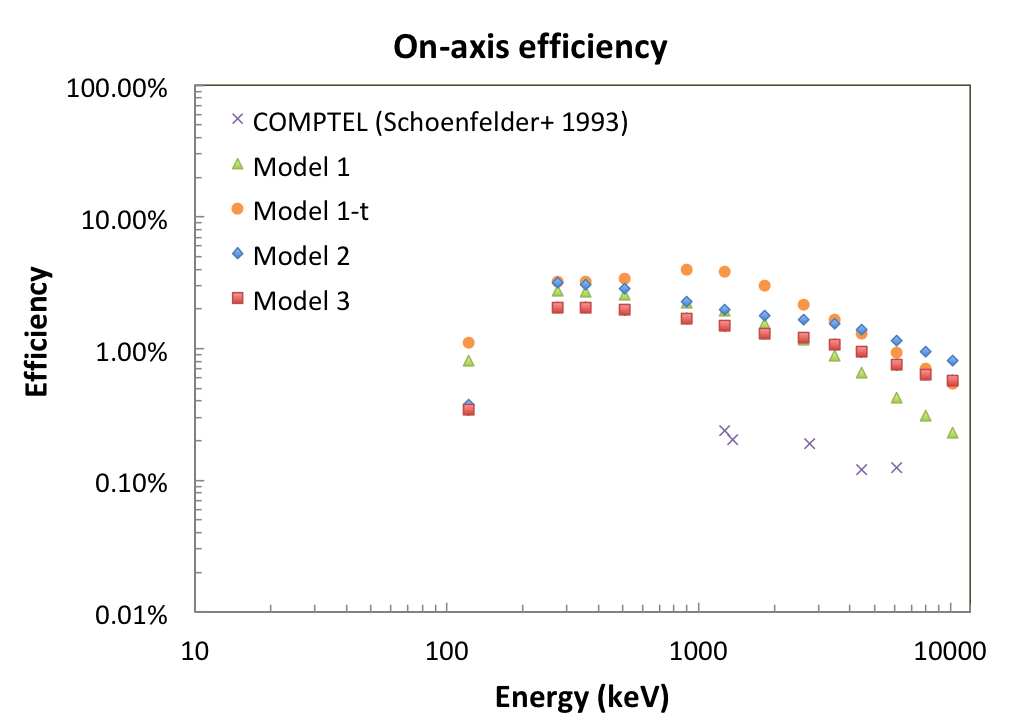}}
\caption{
The on-axis efficiency of the four instrument models. 
The COMPTEL efficiency \citep{schoenfelder93} is plotted for comparison.
We note that the COMPTEL ones are with photo-peak energy cut, while ours are without.
Typically energy cut yields a reduction of a factor of 2 or 3.
}
\label{effi}
\end{figure*}

Efficiency is the ratio of the effective area to the geometric area. 
For the moment we study only the on-axis one.
Since we use only Compton events, that is, those events with multiple hits in the detector volume,
events with all the hits in the same cerium bromide unit will be discarded because those hits cannot be
separated in measurement.
These events, as well as all the single-hit events, can be easily rejected on board to reduce the telemetry load.
These criteria were applied to the study of the down-link data rate, as discussed below.
In computing the Compton efficiency we actually added more selection criteria: we excluded all the Compton events with more than one
hit in any cerium bromide unit; 
for silicon sensors, we rejected events with 3 or more hits in one unit;
all the un-reconstructable multiple-hit events and pair-production events were also excluded.
The efficiency of different models at different energies is shown in Figure \ref{effi}.
\begin{table*}[h]
\caption{Shielding study. We conducted simulations with the background model taken from MEGAlib to compare the down-link data volume with
different shields. All the numbers in the 2nd, 3rd and 4th columns are the number of events normalized to the total trigger number of the instrument without shield (Bare) and are expressed in percentage. 
The instrument used in this simulation was Model 1-t and the simulation exposure time was the time for the bare instrument to have about $6 \times 10^6$ triggers. What we concern most is the 4th column (`data to down link'). The 5th column ('reduction') shows the percentage of reduction
in the data to down link compared with the bare one. The 6th column shows the mass of the shield. The 7th column is simply the ratio of the previous two columns. }
\label{shield}
\begin{center}
\begin{tabular}{lccccccc}
\hline
Shield & Triggered Event & Single-hit Event & Data to D/L & Reduction & Mass & Reduction/Mass \\
 (mm) & (\%) & (\%) & (\%) & (\%) & (kg) & (\%/100g) \\
\hline
\\
Bare    & $100 $ & $86.2$ & $5.39$ & $ -  $ & $ -  $ & $ -  $ \\
Al  (1) & $96.8$ & $83.1$ & $5.38$ & $0.20$ & $0.06$ & $0.33$ \\
Al  (5) & $77.9$ & $64.9$ & $5.08$ & $5.61$ & $0.33$ & $1.70$ \\
Al  (10)& $64.2$ & $51.7$ & $4.98$ & $7.82$ & $0.78$ & $1.00$ \\
CsI (1) & $31.1$ & $17.7$ & $5.24$ & $2.75$ & $0.10$ & $2.75$ \\
CsI (5) & $20.8$ & $8.39$ & $4.89$ & $9.29$ & $0.56$ & $1.66$ \\
CsI (10)& $19.1$ & $6.99$ & $4.84$ & $10.1$ & $1.30$ & $0.78$ \\
Pb  (1) & $20.8$ & $8.26$ & $4.96$ & $7.97$ & $0.24$ & $3.32$ \\
Pb  (5) & $16.6$ & $5.12$ & $4.67$ & $12.9$ & $1.41$ & $0.91$ \\
Pb  (10)& $15.8$ & $4.91$ & $4.59$ & $14.8$ & $3.27$ & $0.45$
\\
\hline
\end{tabular}
\end{center}
\end{table*}
 
In order to find a suitable shielding option to reduce the data amount to down link,
we used the LEO background model incorporated in MEGAlib.
We considered shields made of Al, CsI, and Pb with 1, 5, 10 mm thickness.
The detector model used in this study was Model 1-t.
We compared the shielding effect on the reduction of the data to down link
and also the weight of the shield.
The results are shown in Table \ref{shield}.
For 1-mm Pb shield, the reduction in data to down link is about 8\%, among the high ones, although not the highest.
Its weight, 0.24 kg, is probably still manageable in a 3-U cubesat. 
We chose this shield for the study of down-link data rate for different instrument models.

We used the Cyg X-1 spectrum presented in \citet{laurent11} for 
the estimate of the count rate from the source. Since the flux is low, we had a simulation with a much longer exposure time 
than for estimating the down-link data rate.
The selection criteria of Compton events in estimating the source rate were the same as that in the efficiency study.
For the data rate, single-hit events and those with all the hits in one cerium bromide unit were rejected, as discussed earlier.
The results are shown in Table \ref{rate}. 
The signal-to-noise ratio shown in the last column in Table \ref{rate} is only indicative,
because after event reconstruction one can further reject background events with a spatial cut.
This may reduce the background by a factor of 10 or so. That S/N can be 3 times larger.
We therefore conclude that Model 1-t and Model 2 can achieve significant detection with observation time of several days.
It is quite promising for polarization measurement.  
\begin{table}
\caption{Source count rate and data rate. The data rate is mainly from background counts.
The last column (S/N) is for a 1 Msec exposure and is the ratio of source counts to the square root of background counts. 
As discussed in the main text, 
the background can be further reduced after Compton event reconstruction.}
\label{rate}
\begin{center}
\begin{tabular}{lccccccc}
\hline
Model & Source Rate & Data Rate & S/N    \\
      & (count/sec) & (count/sec) & in $10^6$ sec \\
\hline
\\
  1   &  $0.0043$  &  $2.80$  & 2.5 \\
 1-t  &  $0.0057$  &  $2.15$ & 3.8  \\
 2    &  $0.0045$  &  $1.83$ & 3.2  \\
 3    &  $0.0033$  &  $3.19$  & 1.8 \\
\hline
\end{tabular}
\end{center}
\end{table}

\section{Discussion and Conclusions}

In this paper we concerned ourselves with the design of a small Compton polarimeter to fly on a 3-U cubesat.
We considered low-cost sensors and paid much attention to the down-link demand, which affects the communication requirements
of the cubesat. With the estimate shown in Table \ref{rate}, assuming 20 Bytes for each event, the science data volume per day is about
4 MB only.  Housekeeping data may be a factor of 10 more than that.
Depending on the number of ground stations and the duration of communication contact,  UHF is likely enough.

Based on the above results, Model 1-t and Model 2 seem to be a better design. 
A 1-mm-thick Pb shield also serves better. 
The study on their sensitivity and polarization measurement performance is now on-going.
\vspace{0.3cm}
\\ \noindent 
{\it Acknowledgements}
\vspace{0.3cm}
\\ \noindent 
This work is supported by the Ministry of Science and Technology of the Republic of China (Taiwan) under grant
MOST 107-2119-M-007-012.

\bibliographystyle{aa}

\end{document}